\documentclass[useAMS,usenatbib]{mn2e}

\title{Near-Infrared and Optical Studies of the fast nova V4643 Sgr (Nova Sagittarii 2001)}

\author[N.M. Ashok, D.P.K. Banerjee, W.P. Varricatt $\&$ U.S.Kamath]{N.M. Ashok,$^{1}$
\thanks{E-mail: ashok@prl.ernet.in (NMA); orion@prl.ernet.in (DPKB); 
w.varricatt@jach.hawaii.edu (WPV); kamath@crest.ernet.in (USK)}
D.P.K. Banerjee$^{1}$, W.P. Varricatt$^{2}$, U.S. Kamath$^{3}$\\
$^{1}$Physical Research Laboratory, Navrangpura, Ahmedabad 380009, India\\
$^{2}$Joint Astronomy Center, 660 North A'ohoku Place, Hilo, HI 96720, USA\\
$^{3}$CREST, Indian Institute of Astrophysics, Bangalore 560034,India}

\usepackage{graphicx}
\begin{document}

\date{Accepted  Received }

\pagerange{\pageref{firstpage}--\pageref{lastpage}} \pubyear{2002}

\maketitle

\label{firstpage}

\begin{abstract}
V4643 Sagittarii or  Nova Sagittarii 2001 was discovered in outburst at 7.7 mag. on 2001 February 24. Here, we present near-infrared results of this fast classical nova  obtained in the early decline phase in 2001 March followed by optical observations about one month later. Subsequently we also present  near-infrared spectra taken later in the nova's evolution, about four months after the outburst, when V4643 Sgr had entered the coronal phase. The spectra in the early decline phase are  dominated by emission lines of the H \,{\sc i} Brackett series and also the Paschen $\beta$ and $\gamma$ lines. We study the cause of the excitation of the the O\,{\sc i} line at 1.128 ${\rm{\mu}}$m and discuss the  variation in its strength with time after outburst. We discuss the role of optical depth effects on the observed strengths of the hydrogen Brackett and Paschen lines and discuss possible reasons for  the puzzling behavior of the Br $\gamma$ line strength and whether it is correlated with the O\,{\sc i} 1.128 ${\rm{\mu}}$m line behavior.  An optical spectrum is presented which shows that  He \,{\sc ii} lines are the most prominent features - after H \,{\sc i} - to be seen in early 2001 April.  We present and also discuss spectra taken in 2001 June and August which prominently show coronal lines of [Si \,{\sc vi}] and [Si \,{\sc vii}] at 1.9641 ${\rm{\mu}}$m and  2.4807 ${\rm{\mu}}$m  respectively.

\end{abstract}


\section{Introduction}
V4643 Sagittarii or  Nova Sagittarii 2001 was discovered in outburst at 7.7 mag by Liller (2001) on 2001 Feb 24.369. Nothing brighter than  a limiting magnitude
of  11.1  appeared at the position of the nova on photographs taken by
Nakamura (2001) on 2001 Feb 20.852. This indicates that the nova was
detected close to  maximum light. Samus  (2001) has identified the possible 
progenitor with a star in the USNO-A2.0 catalog (identification number 0600-29446361) which lies within 1 arc sec of the nova's position having blue and red
magnitudes of 17.4 and 15.8 respectively. Initial optical spectra of
the nova by Della Valle et al. (2001) on 2001 Feb 26.35 UT show it to
be dominated by broad emission lines of the hydrogen Balmer series. Other
lines  due to  Na \,{\sc i}, He \,{\sc i}, N \,{\sc ii}, He \,{\sc ii} and O\,{\sc i} were also seen in the
spectra. Della Valle et al. (2001) point out that the absence of Fe \,{\sc ii} lines
and the large full width at zero intensity (FWZI) of the H$\alpha$ profile, 10000 km s${^{\rm -1}}$, implies that V4643 Sgr belongs to the He/N class - also referred to as the ONeMg class -  of novae (Williams 1992). The detailed light curve of the object covering the initial 50 days is shown in Bruch (2001). Assuming that V4643 Sgr was detected very close to its outburst maximum, Bruch (2001) has estimated a 
t$_{\rm 2}$ and  t$_{\rm 3}$ of 4.8 and 8.6 days respectively (where t$_{\rm 2}$ and  t$_{\rm 3}$ are the time taken to fall by 2  and 3 magnitudes from maximum). This shows that V4643 Sgr belongs to the group of fast novae like V1500 Cyg, V838 Her and V2487 Oph.\\

 In this work we present near-IR $JHK$ spectra of V4643 Sgr at four
different epochs. The spectra essentially cover two phases of the
nova's evolution viz. i) spectra obtained shortly after the outburst 
when it was dominated by hydrogen emission lines (Ashok, Tej $\&$ Banerjee, 2001) and ii) spectra obtained when the nova had entered the coronal phase (Ashok, Banerjee $\&$ Varricatt, 2001). In addition we also present an optical spectrum taken about a month after the outburst. Based on these observations, we discuss the properties and temporal evolution of V4643 Sgr. \\

\section{Observations}
Near-IR $JHK$ spectra were obtained at the Mt. Abu 1.2m telescope in the
early decline phase during 2001 March and later during 2001 June and August
from the United Kingdom Infrared Telescope (UKIRT). The log of the observations 
is given in Table 1.\\

      The Mt. Abu spectra were obtained at a resolution of $\sim$ 1000
using a Near-Infrared Imager/Spectrometer with a 256$\times$256 HgCdTe
NICMOS3 array. In each of the $JHK$ bands a set of spectra were taken with the
nova off-set to two different positions along the slit ( slit width 1 arc 
second). Spectral calibration   was done using the OH sky lines 
that register with the stellar spectra. The spectra of the comparison star
HR 6490 ( for 2001 March 2 and 3) and HR 6486 (for 2001 March 14, 15 and 16) were
taken at similar airmass as that of V4643 Sgr to ensure that the
ratioing process (nova spectrum divided by the standard star spectrum)  removes the telluric lines reliably.  To avoid artificially generated emission lines in the ratioed
spectrum - due to HI lines in the spectrum of the standard star - the hydrogen absorption lines in the spectra of the standard star were removed by interpolation before ratioing. The ratioed spectra were then multiplied by a blackbody curve
corresponding to the standard star's effective tempearture to yield the final spectra.
The UKIRT spectra were obtained with the Cooled Grating Spectrograph (CGS4)
using the 40l/mm grating with the J band spectra being taken in the second order and the H, K band spectra taken in the first order. The UKIRT spectra were wavelength calibrated using arc spectra.\\

\begin{table}
\caption{A log of the spectroscopic observations of V4643 Sgr. The date of 
outburst has been assumed to be its detection date viz. 2001 Feb 24.369 UT}
\begin{tabular}{llllll}
\hline \\ 
Date & Days & Telescope & Spectral & Inte-\\
2001 & since &  & Band & gration\\
(UT) & Outburst &  &  & time (s)\\
\hline 
\hline \\ 
Near Infrared &     & & &                    \\
& & & & \\
Mar. 2.025  & 5   & Mt. Abu & J	 	& 240\\
            &     &  "      & H         & 120\\
            &     &  "      & K         & 120\\
Mar. 3.021  & 6   &  "      & J    	& 240\\
            &     &  "      & H	        & 240\\
            &     &  "      & K         & 240\\         
Mar. 14.031 & 17  & Mt. Abu & J	        & 120\\
            &     &  "      & H         & 120\\
            &     &  "      & K         & 120\\
Mar. 15.00  & 18  &  "      & J         & 120\\
            &     &  "      & H         & 120\\
            &     &  "      & K         & 120\\
Mar. 15.979 & 19  &  "      & J         & 240\\
            &     &  "      & H         & 240\\
            &     &  "      & K         & 240\\                    
June 16.429 & 112 & UKIRT   & K         & 240\\
June 28.520 & 124 & "       & J         & 240\\
June 29.456 & 125 & "       & H         & 240\\
Aug. 12.326 & 138 & "       & J         & 1200\\
Aug. 12.349 &  "  & "       & H         & 960\\
Aug. 12.299 &  "  & "       & K         & 960\\
& & & &  \\
Optical & & & & \\
April 1   &  36 & VBT     & Visible   & 600\\ 
                                                         
\hline
\end{tabular} 
\end{table} 
Photometry in the $JHK$ bands was also done from Mt. Abu
on 2001 May 11 in photometric sky conditions using the NICMOS3 array in the imaging mode.
Several frames,  in 4 dithered positions, offset by 
30 arcsec were obtained in all the filters with exposure times for the individual J, H, K frames being 60, 20 and 0.5 s respectively. The sky frames, which are subtracted from the nova frames, were generated by median combining the dithered frames. The standard star HD 161903 was used for photometric calibration  on 2001 May 11.  We also derived JHK magnitudes from the UKIRT spectroscopic observations of 2001 August 12 -  the sky being photometric on this particular night.  The results from the photometry are presented  in Table 4 and discussed later.\\ 
 
An optical spectrum (4600-8200 \AA, 2-pixel resolution of 11 \AA)  was obtained using an OMR spectrograph at the Cassegrain focus of the 2.3m Vainu Bappu Telescope, Kavalur on 2001 April 1, 36 days after discovery. Feige 34 was used as the standard star. Wavelength calibration was established using FeAr spectrum. The data - both infrared and optical - were reduced 
and analyzed using IRAF and Starlink packages. 

\begin{figure}
\centering
\includegraphics[bb= 67 247 490 609, width=3.2in,height=2.8in,clip]{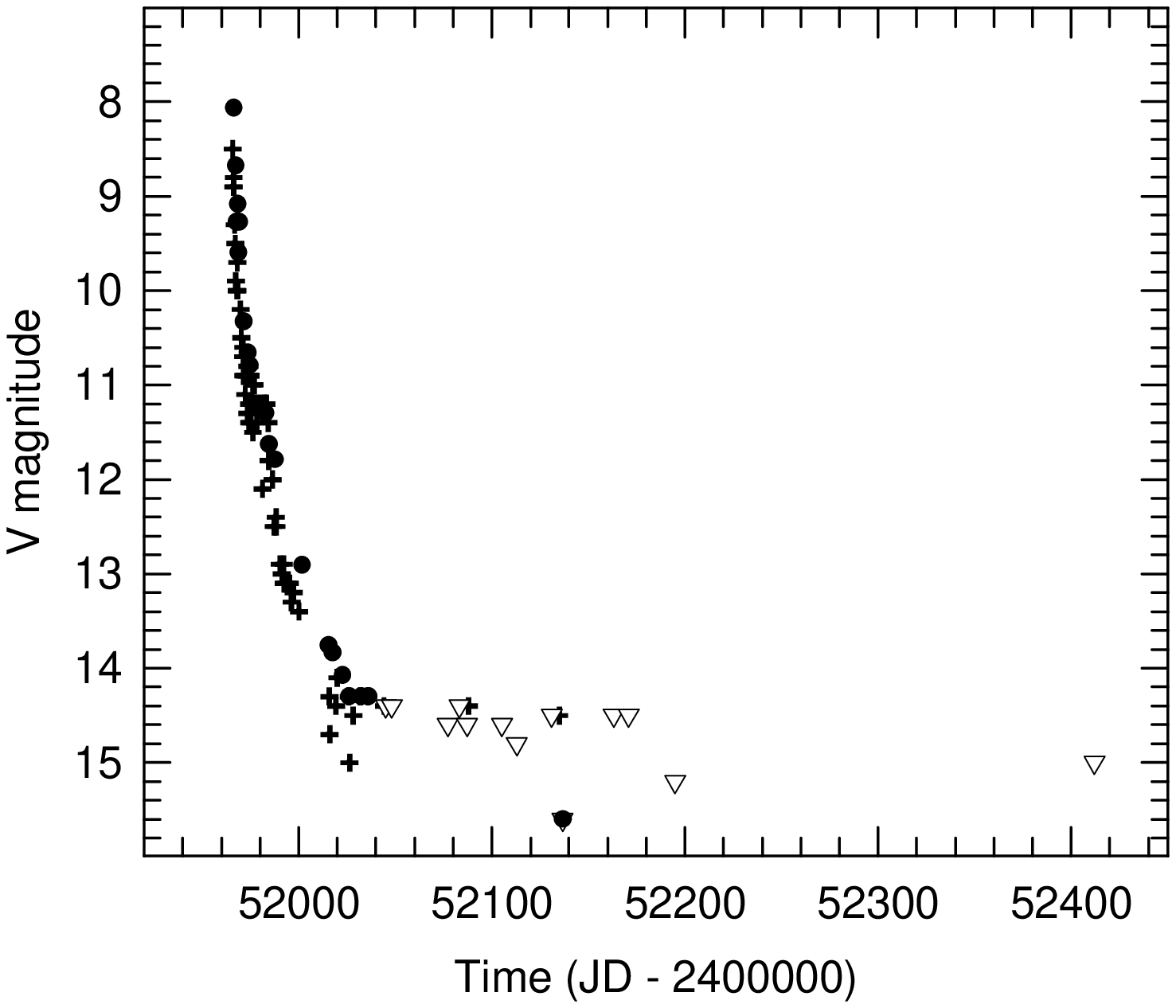}
\caption[]{ The visual light curve of V4643 Sgr. Crosses and open, inverted triangles are visual estimates from data from AFOEV - the majority of these observations coming from the Variable Star Section of the Royal Astronomical Society of New Zealand. The triangles represent  limiting magnitudes. The filled circles are data from the Variable Stars Network (VSNET), Japan.}
\label{fig1}
\end{figure}


\begin{figure}
\centering
\includegraphics[bb=113 123 396  541,width=3.2in,height=4.5in,clip]{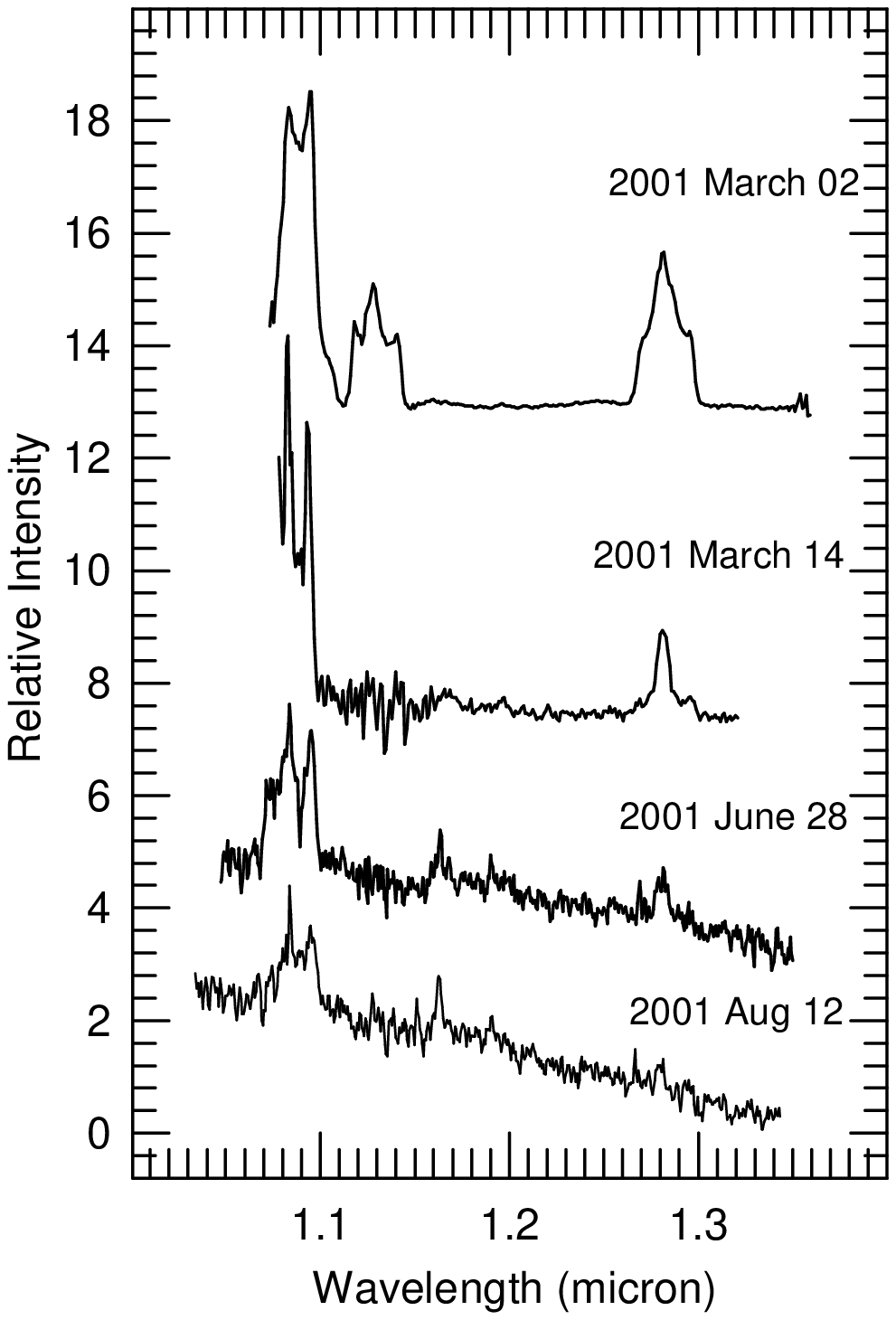}
\caption[]{ The $J$ band spectra of V4643 Sgr are shown at different epochs.
The spectra have been offset from each other for clarity.}
\label{fig2}
\end{figure}


\begin{figure}
\centering
\includegraphics[bb=112 126 392 544,width=3.2in,height=4.5in,clip]{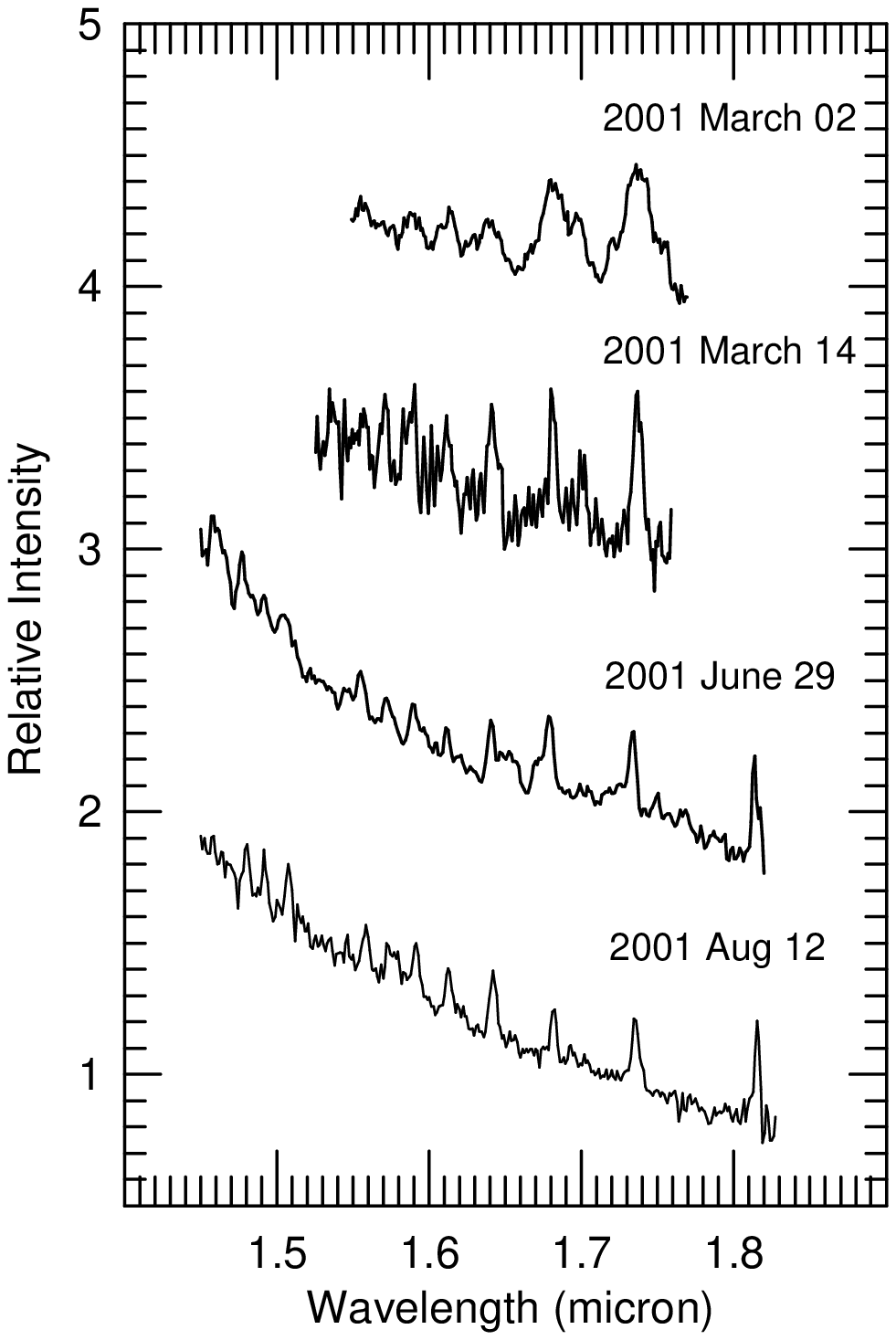}
\caption[]{The $H$ band spectra of V4643 Sgr are shown at different epochs. 
The spectra have been offset from each other for clarity.}
\label{fig3}
\end{figure}

\section{Results}
Before presenting the results proper, we estimate some of the  useful
parameters of V4643 Sgr.

\subsection{Light curve, Outburst luminosity, reddening and distance} We present the light curve of the object in Figure 1 using  additional data beyond that presented in the light curve by Bruch (2001). The fast decline of the nova in the early stage may be noted. The absolute magnitudeof V4643 Sgr has been estimated by Bruch (2001) using the different MMRD
relations ( maximum magnitude - rate of decline) available in the literature.
A consistent value of $M{_{\rm V}}$ = -9.04 $\pm$ 0.08 is found from the different MMRD relations. We derive the reddening from the method of van den Bergh $\&$ Younger (1987) who show that at t$_{\rm 2}$,  the $(B-V)$ colors of novae are $(B-V)$ = -0.02 $\pm$ 0.04. Taking the  discovery epoch 2001 Feb 24.369 as the time of optical maximum and t$_{\rm 2}$ = 4.8 days (Bruch 2001) we have used data from VSNET to calculate $E(B-V)$ at t$_{\rm 2}$ (i.e. on 2001 March 1.169) . The optical photometry reported in VSNET adjacent to 2001 March 01 is given in Table 2. Interpolating the $(B-V)$ colors given in Table 2, we get $(B-V)$ = 1.45 at t$_{\rm 2}$, resulting in $E(B-V)$ = 1.47 and $A{_{\rm V}}$ = 4.56 for $R$ = 3.1. Such a large value for $A{_{\rm V}}$ is expected as the nova is located close to the galactic plane with $b{^{\rm II}}$ = -0.34. The distance to the nova is calculated using the standard relation 
$m{_{\rm V}}$ - $M{_{\rm V}}$ = 5log$d$ - 5 + $A{_{\rm V}}$.
For $M{_{\rm V}}$ = -9.04, $m{_{\rm V}}$ = 8.1 and $A{_{\rm V}}$ = 4.56, the distance is estimated to be 3.3 kpc. 

\begin{table}
\caption{$BV$ photometry data from VSNET}
\begin{tabular}{lllllll}
\hline \\ 
Date 2001 & B & V & $(B-V)$&\\ 
(UT) & &  &  & \\
\hline 
\hline \\ 
February 26.817  & 10.87   &  9.28  & 1.59 & \\
February 27.835  & 11.08   &  9.60  & 1.48 & \\
March     2.183  & 11.75   & 10.33  & 1.42 & \\

\hline
\end{tabular} 
\end{table}

\begin{figure}
\centering
\includegraphics[bb=118 118 406 539,width=3.2in,height=4.5in,clip]{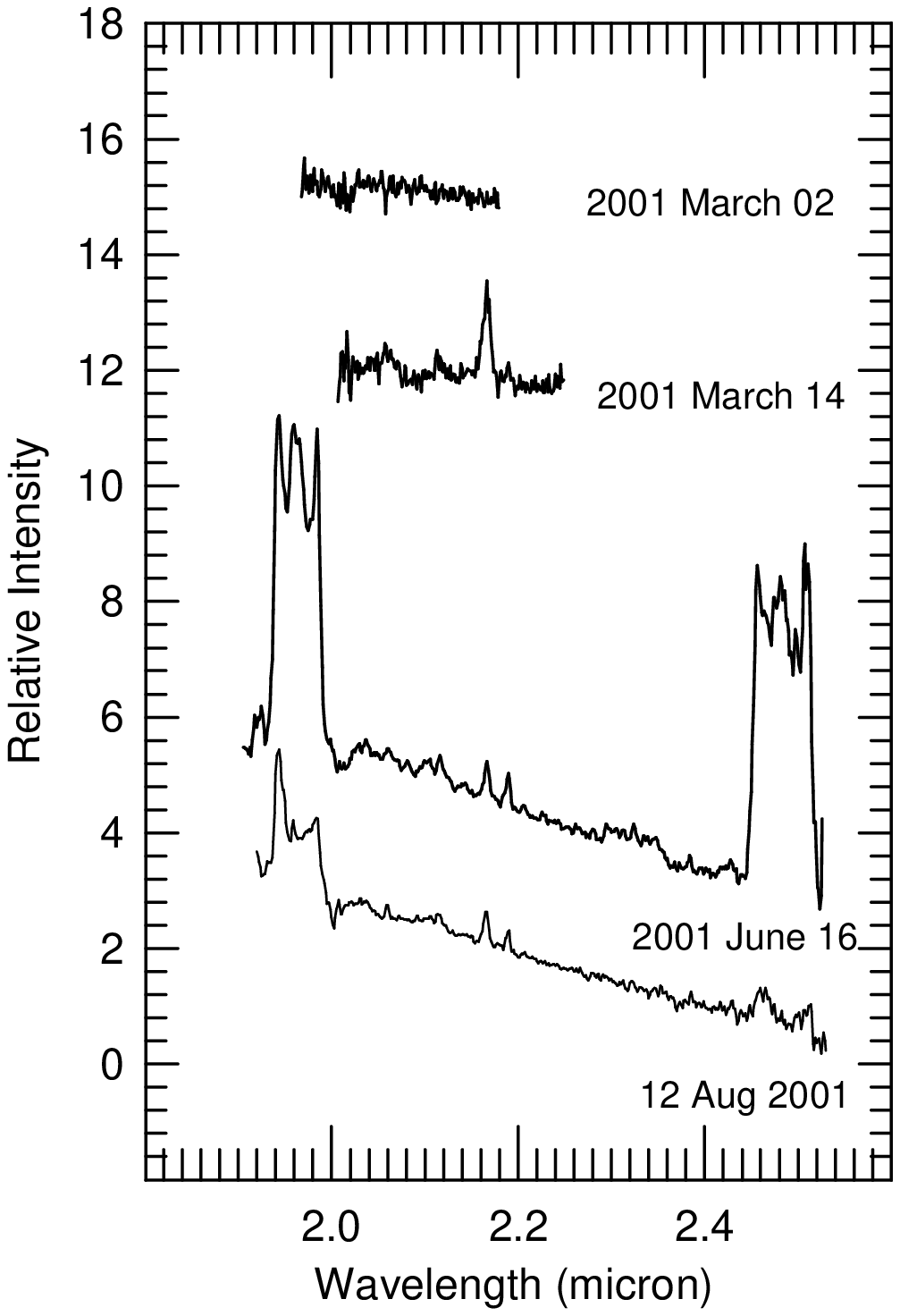}
\caption[]{The $K$ band spectra of V4643 Sgr are shown at different epochs. 
The spectra have been offset from each other for clarity.}
\label{fig4}
\end{figure}

\subsection{$JHK$ spectroscopy}The $JHK$ spectra at different epochs are
presented in Figures 2, 3 and 4. The prominent lines seen in the early decline
phase are the hydrogen Brackett and Paschen series lines and the 
O\,{\sc i} 1.128 ${\rm{\mu}}$m line. Later in the evolution, during the coronal phase, 
the [Si \,{\sc vi}] at 1.9641 ${\rm{\mu}}$m and [Si \,{\sc vii}] at 2.4807 ${\rm{\mu}}$m lines are the most prominent emission features in the observed spectra. The details of the line identification are presented in Table 3 - some unidentified lines often seen in novae spectra are marked as u.i. Also given in Table 3 are  equivalent widths (W) on selected days viz. 2005 March 2 $\&$ 14 and June 12. The W values on the first two of these days are representative for subsequent discussions in Sections 3.3 and 3.4 regarding the strength of the H \,{\sc i} lines relative to themselves or to the O\,{\sc i} lines. The W values of June 12 are specially relevant in context of the the Si coronal lines , seen  most prominently at this epoch,  and whose relative strengths are used subsequently in Section 3.7 to derive the coronal temperature. We discuss the 
spectral lines and their evolution in greater details in the coming sub-sections.

\begin{table}
\caption[]{List of observed lines in the JHK spectra}
\begin{tabular}{llllll}
\hline\\
Wavelength & Species  & Eq.W & Eq.W & Eq.W \\
(${\rm{\mu}}$m) & & (\AA) & (\AA) & (\AA) \\
& & March 2 & March 14 &  June 16\\
\hline 
\hline \\ 
1.0830   & He \,{\sc ii}      & He \,{\sc i} 1.08    &          &     \\
1.0938   & Pa $\gamma$        & $\&$ Pa $\gamma$ are &          &     \\
         &                    &  blended             &          &     \\
1.1287   & O\,{\sc i}         & 435                  & absent   &     \\
1.1626   & He \,{\sc ii}      &                      &          &     \\
1.1900   & u.i                &                      &          &     \\
1.2818   & Pa $\beta$         & 568                  &  72      &     \\
1.3164   & O\,{\sc i}         &                      &          &     \\
1.4760   & He \,{\sc ii}      &                      &          &     \\
1.4882   & He \,{\sc i}       &                      &          &     \\
1.5439   & Br 17              &                      &          &     \\
1.5557   & Br 16              & 5                    &  5.4     &     \\
1.5701   & Br 15              &                      &  12      &     \\
1.5881   & Br 14              & 14                   &  29      &     \\
1.6109   & Br 13              & 16                   &  12      &     \\
1.6407   & Br 12              & 22                   &  31      &     \\
1.6806   & Br 11              & 94                   &  17      &     \\
1.7002   & He \,{\sc i}       &                      &          &     \\
1.7362   & Br 10              & 128                  &  43      &     \\
1.8174   & Br  9              &                      &          &     \\
1.9641   & [Si \,{\sc vi}]    &                      &          & 426 \\
2.100    &  u.i               &                      &          &     \\
2.1120   &  He \,{\sc i}      &                      &          &     \\
2.1132   &  He \,{\sc i}      &                      &          &     \\
2.1655   & Br $\gamma$        & absent               & 75.4     &     \\
2.1882   & He \,{\sc ii}      &                      &          &     \\
2.3205   & [Ca\,{\sc viii}]   &                      &          & 43  \\
2.4807   & [Si \,{\sc vii}]   &                      &          & 896 \\

\hline
\end{tabular} 
\end{table}

\subsection{ The behavior of the O\,{\sc i} 1.128 ${\rm{\mu}}$m line} 
The O\,{\sc i} 1.128 ${\rm{\mu}}$m line is very prominently seen in the J
band spectra of 2001 March 2 whereas the O\,{\sc i} 1.316 ${\rm{\mu}}$m line is absent.
This implies that continuum fluorescence cannot be the source of
excitation of these O\,{\sc i} lines because the predicted strength of the
lines is W(1.3164)/W(1.1287) $\geq$ 1 if continuum fluorescence is the
significant excitation mechanism (Strittmatter et al. 1977 and
references therein; Grandi 1980). The large strength
of the O\,{\sc i} 1.128 ${\rm{\mu}}$m line indicates that Lyman (Ly) $\beta$ fluorescence is the pumping mechanism. As is known, due to the  close matching of energy levels,  Ly $\beta$ photons can pump the O\,{\sc i} ground state resonance line at 1025.77A. The generally accepted mechanism for Ly $\beta$ fluorescence to operate has been explained by Grandi (1980). This mechanism, used to explain the O\,{\sc i} emission in Seyfert galaxies, is also applicable to nova shells (for e.g. Strittmatter et al. 1977). In this scenario  it is necessary to have a large population of neutral O\,{\sc i}, a source of Ly$\beta$ photons and a large optical depth in H$\alpha$. Ionizing photons beyond the Ly continuum will not be able to penetrate into the deep, interior regions of the nova shell. At such sites hydrogen and oxygen can remain neutral as  both species have similar ionization potentials of 13.6 and 13.62 eV respectively. However, Ly$\alpha$ photons, formed at the ionized outer regions of the shell can migrate to the inner regions of the shell and get trapped. These 
Ly$\alpha$ photons can subsequently excite the neutral hydrogen to the n=2 level, thereby creating the requirement of a large optical depth in H$\alpha$. Although, matter deep within the nova shell is shielded from Ly continuum photons from the central photoionizing  source, Balmer continuum photons can penetrate to such sites  and ionize the considerable population of hydrogen atoms in the n=2 levels. Recombinations from such photoionizations, followed by downward cascading, can then produce the  Ly$\beta$ photons needed to excite the O\,{\sc i} line.
 
In the case of V4643 Sgr it is necessary to see whether there is evidence for a 
large optical depth in H$\alpha$. We follow the analysis, done by Strittmatter et al. (1977) for Nova Cyg 1975 (V1500 Cygni), in predicting the relative strengths of the O\,{\sc i} 
and H$\alpha$ lines.  They show that, if Ly$\beta$ fluorescence is the pumping mechanism, then the ratio I (H$\alpha$)/I (O\,{\sc i} 8446) should be very large ($\sim$ 7500) where 
I (H$\alpha$) and I (O\,{\sc i} 8446) are the intensities of the 
H$\alpha$ and the O\,{\sc i} 8446 line. This assumes a normal  abundance for oxygen, but as is known, the oxygen mass-fraction in the ejecta of ONeMg novae like V4643 Sgr can be enhanced by a factor of 6 to 7 (e.g Starrfield et al.,  1997). But even after taking this into account,  a large value of the I (H$\alpha$)/I (O\,{\sc i} 8446) ratio is expected. It may be noted that other processes like recombination can add to the strength of the H$\alpha$ line thereby increasing this ratio. On the other hand,  the O\,{\sc i} 8446 line, as can be seen from the energy level diagram for O\,{\sc i} ( e.g. Figure 3 in Grandi 1980), is fed only by the O\,{\sc i} 1.128 ${\rm{\mu}}$m line in the case of pure Ly$\beta$ fluorescence. Hence the number of photons in the 8446\AA\ and 1.128 ${\rm{\mu}}$m lines are expected to be the same (Venturini et al. 2002). Thus the ratio of I (H$\alpha$) / I (O\,{\sc i} 1.128) is also predicted to be very large. The observed equivalent widths, that are available for the 
H$\alpha$ line, are 450\AA\ on 2001 February 25.88 (Ayani $\&$ Kawabata, 2001), 21.9\AA\ and 6.8\AA\ on 2001 March 16 and 2001 May 4 respectively (Bruch, 2001). The first of these values can be compared with the equivalent width of the  O\,{\sc i} 1.128 ${\rm{\mu}}$m line on  March 2 
(eq. width = 435\AA\ ), the observation epochs being quite close. As can be seen, the ratio of the equivalent widths is closer to 1 and departs radically from the predicted value of 
I (H$\alpha$)/ I (O\,{\sc i} 1.128) thereby indicating a high optical depth in the  
H$\alpha$ line. Thus the requirement of a large optical depth in H$\alpha$ for effective 
Ly$\beta$ pumping of the O\,{\sc i} line is satisfied. In fact, as discussed in the next subsection, there seems to be large optical depth effects not only in the H$\alpha$ line but also in the hydrogen lines of the Paschen and Brackett series.
     
\subsection{ The near-infrared Hydrogen lines}
Only two of the  Paschen series lines are covered in the spectra presented here viz.  
Pa $\beta$ at 1.2818 ${\rm{\mu}}$m and Pa $\gamma$ at 1.0938 ${\rm{\mu}}$m. Both these Paschen lines are seen prominently  soon after the outburst and then decline in intensity with time. Since Pa $\gamma$ is strongly blended with the He\,{\sc i} at 1.0830 ${\rm{\mu}}$m, it is difficult to estimate its equivalent width and compare its strength with Pa $\beta$. However, even a visual inspection suggests Pa $\gamma$ to be stronger (or comparable) than Pa $\beta$ at all epochs of our observations. This indicates that the Paschen series lines are optically thick since the  expected ratio in recombination case B conditions 
is I(Pa $\beta$)/I(Pa $\gamma$) $\sim$ 1.6 (for $T$ $=$ 1 $\times$ 10${^{\rm 4}}$ K,  $n$${_{\rm e}}$ $=$ 6$\times$10${^{\rm 10}}$ cm${^{\rm -3}}$ ). It is also pertinent to note that Pa $\beta$, Pa $\gamma$ and the O\,{\sc i} 1.128 ${\rm{\mu}}$m lines have broad wings with a relatively narrow peak at the center - a profile similar to that seen in V2487 Oph (Nova Oph 1998; Lynch et al. 2000)
  
Optical depth effects are more clearly seen in the Brackett series lines in the 
$H$ band. In the 2001 March spectra, while the higher Br series lines are clearly seen in the $H$ band, Br $\gamma$ which is expected to be much stronger in comparison is completely absent (Figure 4). This is a puzzling result. By March 14, the 
Br $\gamma$ line  has begun to appear in the $K$ band spectra and strangely this coincides with the disappearance of the O\,{\sc i} 1.128 ${\rm{\mu}}$m line. We discuss the possible significance of this coincidence shortly.  In Figure 5, we present  plots of  the observed strength of the Brackett lines versus their predicted intensities in a  recombination case B condition. The case B line intensities are from Storey $\&$ Hummer (1995) and  assume a temperature $T$ $=$ 1 $\times$ 10${^{\rm 4}}$ K and electron density 
$n$${_{\rm e}}$ $=$ 6$\times$10${^{\rm 10}}$ cm${^{\rm -3}}$. For the data of 2001 March 2 and 3, since Br $\gamma$ was not detected at all, we have assumed its equivalent width to be zero. At most it can be less than (or equal to) that associated with modulations in the continuum due to noise. We have also deblended the Br11 line at 1.6807 ${\rm{\mu}}$m from the nearby 
He\,{\sc i} line at 1.7002 ${\rm{\mu}}$m in estimating the former's equivalent width. As can be seen from all three panels of Figure 5, the observed line intensities  deviate significantly from the optically thin case B values. This indicates that Brackett lines  are optically thick both during the early decline phase and also at later stages.\\  
    
      Similar optical depth effects, in the near-IR hydrogen lines, were observed by Lynch et al. (2000) in the fast Nova Oph 1998(V2487 Oph). These authors have developed a model to explain the observed line strengths and have shown that the relatively larger intensities of the higher members of the Paschen and Brackett series arise because of emission from high-density or optically thick emission-line gas. Increased strength in the  higher lines occurs when the level populations become thermalized at high densities ( $n$${_{\rm e}}$ $\geq$ 10$\times$10${^{\rm 10}}$ cm${^{\rm -3}}$) or at large optical depths. In such cases radiative decays become less important relative to electron collisions in determining level populations. We show a sample model fit to the data of 2001 August 12 (bottom panel, Fig. 5), using the tabulated results given in Lynch et al. (2000). The model values are the expected line strengths for a gas having
$n$${_{\rm e}}$ $=$ 6$\times$10${^{\rm 11}}$ cm${^{\rm -3}}$,
$T$ $=$ 1.0 $\times$ 10${^{\rm 4}}$ K and an optical depth 
$\tau$ $=$ 100 ($\tau$ has been given at the
Pa ${\rm{\alpha}}$ line-center).  We tried model fits for other combinations of the parameters $n$${_{\rm e}}$, $T$ and $\tau$ since Lynch et al. (2000) give expected line strength data for combinations of $n$${_{\rm e}}$ $=$ 6$\times$10${^{\rm 10}}$ 
and 6$\times$10${^{\rm 11}}$ cm${^{\rm -3}}$; 
$T$ $=$ 5$\times$10${^{\rm 3}}$, 1.0$\times$10${^{\rm 4}}$, 
1.5$\times$10${^{\rm 4}}$ K; 
and ${\rm{\tau}}$ $=$ 10, 100 and 1000.  However, we find that the excessive strength of the higher lines vis-a-vis the lower lines like Br${\rm{\gamma}}$ (just the opposite of Case B predictions) can only be explained by invoking large optical depth values for the parameter 
${\rm{\tau}}$. It may be pointed out that the data for 2001 March 14-16 (middle panel, Fig. 5) is also reasonably well fitted by Lynch et al. (2000) model fits (with $n$${_{\rm e}}$ $=$ 6$\times$10${^{\rm 11}}$ cm${^{\rm -3}}$, $T$ $=$ 5$\times$ 10${^{\rm 3}}$ K, $\tau$ $=$ 100) but not the 22001 March 2 data where the optical depth in the Br $\gamma$ line is most pronounced - in fact the line is absent. This could possibly be accounted for by extending the model calculations for even larger optical depths than listed in their work.

The lack of spectral data between 2001 March 2 to 13 does not permit us to
conclude when exactly the O\,{\sc i} 1.128 ${\rm{\mu}}$m line disappeared and the Br $\gamma$ line appeared. But the quasi-simultaneity of  both phenomena indicate that they may be
correlated and its implications therefore warrant atleast a qualitative explanation. If it is accepted that O\,{\sc i} is caused by Ly $\beta$ fluorescence then its fading would imply a lack of Ly $\beta$  photons for excitation. If this mechanism - described in detail earlier-  is reviewed, such a possibility is unlikely. In this scenario the absence of Ly $\beta$ photons should also lead to the absence of Paschen and Brackett series photons since all these quanta are produced by recombination and downward cascading. Since this is not what is observed, we conjecture that the disappearance of O\,{\sc i} is due to its destruction by photo-ionization by a small fraction of Ly continuum photons from the central star. Such photons, which earlier could not penetrate into the core of the nova shell, due to large optical depth in the Ly continuum, are more likely to do so after the rapid expansion of the nova.   We note that the absence of the  O\,{\sc i} 1.128 ${\rm{\mu}}$m line in V4633 Sgr 525 days after the outburst was similarly interpreted by Lynch et al. (2001) to infer a considerable  optical thinning of the ejecta.
In the absence of frequent near-infrared spectroscopic monitoring of novae in the early decline phase it is difficult to determine when the O\,{\sc i} 1.128 ${\rm{\mu}}$m line disappears but the present observations suggest that in a fast nova this could occur within a rather short period of $\sim$ 10 days. A general detailed study on the variation of the OI line strength with time, based on all novae spectra in which the line has been detected, is intended to be pursued as a part of a separate work.

\begin{figure}
\centering
\includegraphics[bb=80 50 466 656, width=3.2in,height=4.5in,clip]{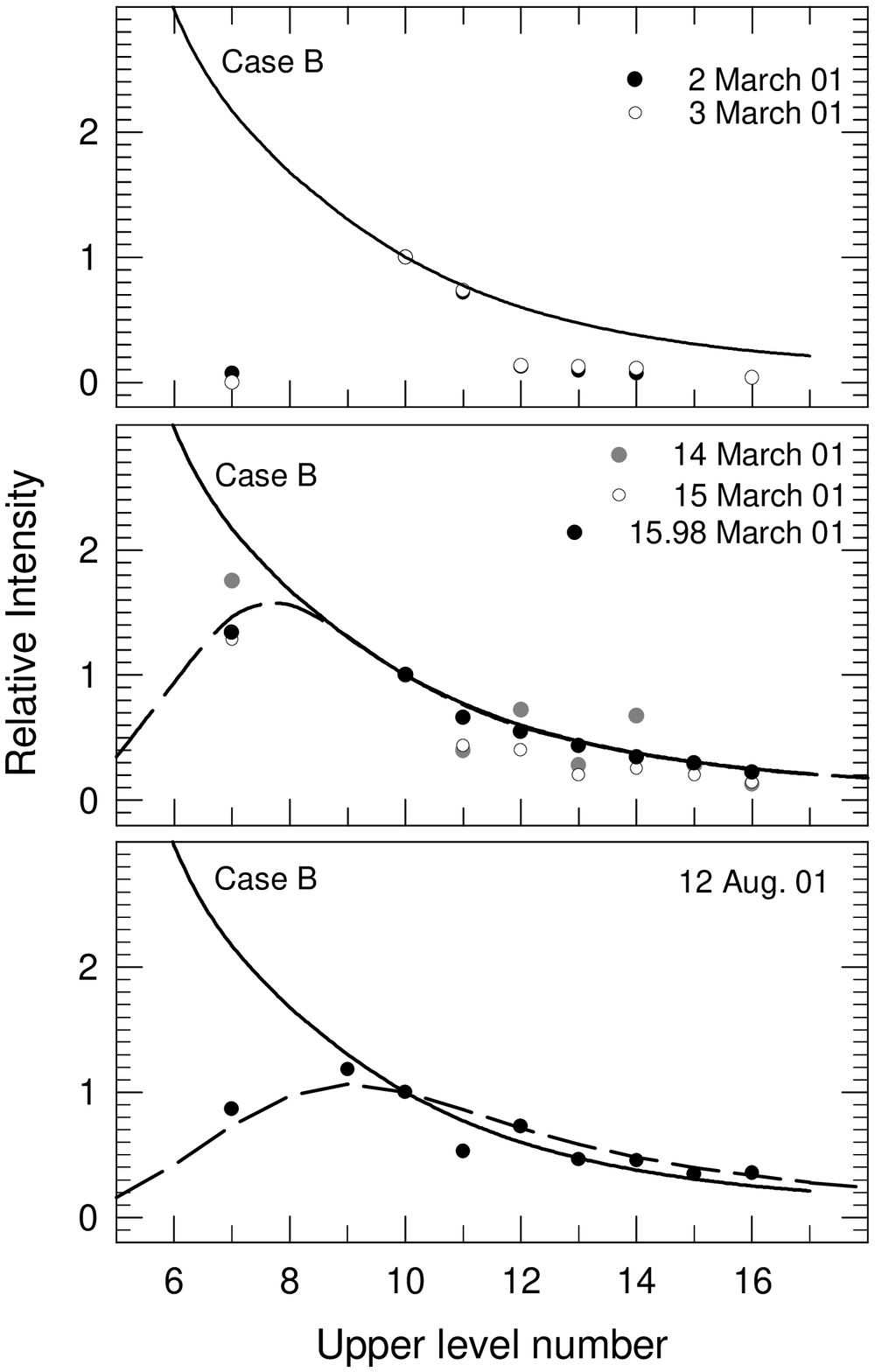}
\caption[]{Optical depth effects seen in the hydrogen Brackett lines in V4643 Sgr. The abscissa gives the upper level number of the Brackett line transition. In all three panels, the Case B line strengths are shown by the continuous line and observed line strengths on different days are marked with black, gray or empty circles. The strength of Br10 line has been normalised to unity. Model fits to 
observed values, based on 
data from Lynch et al. 2001, are shown by the broken lines. The Brackett $\gamma$ line is generally found to be weaker than expected (vis-a-vis the Case B predictions) than the higher lines in the series indicating that
the H\,{\sc i} lines are optically thick. }
\label{fig5}
\end{figure}

\subsection{The evolution of the infrared continuum}
 
We used our photometric data, though not extensive, to study the shape of near-IR continuum.
The JHK magnitudes of Table 4 were corrected for interstellar extinction using  Koornneef's (1983) relations viz. 
$A$${_{\rm V}}$ $=$ 3.1$E(B-V)$, 
$A$${_{\rm J}}$ $=$ 0.265 $A$${_{\rm V}}$,
$A$${_{\rm H}}$ $=$ 0.155 $A$${_{\rm V}}$ and
$A$${_{\rm K}}$ $=$ 0.090 $A$${_{\rm V}}$ and flux calibrated by using zero magnitude
fluxes from Koornneef (1983). We have adopted a value of $A{_{\rm V}}$ = 4.56 (as discussed in section 3.1) in our calculations. After plotting the spectral energy distribution using the JHK data, we do not see any evidence for infrared excess indicating the absence of dust formation in this nova at these epochs. Among the fast novae, only V838 Her formed optically thin dust soon after its outburst (Chandrasekhar, Ashok $\&$ Ragland, 1992). The typical time scales for dust formation in novae, where it has been observed to form, are 50 to 70 days after outburst (Gehrz 1988). We also find from the data of Table 4 that the shape of the near-IR continuum (F${_{\lambda}}$ versus $\lambda$) shows a trend of becoming less steep with time. On 2001 May 11, the continuum shows a $\lambda$${^{\rm -3}}$ dependence while on 2001 August 12 it is closer to $\lambda$${^{\rm -2}}$. In 
general, the evolution of the continuum of novae is not too clearly understood and shows 
considerable diversity in different novae (e.g. Ennis et al.  1977; Lynch et al. 2000, 2001).

\begin{table}
\caption[]{$JHK$ photometry of V4643 Sagittarii}
\begin{tabular}{llll}
\hline \\
Obs. date (UT)&  $J$ & $H$  & $K$ \\
\hline 
\hline \\ 
2001 May 11& 11.3  $\pm$ 0.1   & 10.86 $\pm$ 0.1  & 10.80 $\pm$  0.15  \\
2001 Aug 12& 13.57 $\pm$ 0.1   & 12.75 $\pm$ 0.1  & 12.40 $\pm$  0.1\\

\hline
\end{tabular} 
\end{table}

\subsection{Optical observations}
We could obtain only one optical spectrum of the nova on 2001 April 1 which is shown 
in Figure 6. The spectrum is typical of a He/N nova.
The detailed H${\alpha}$  line profile, shown in Figure 7 , consists of a narrow component surrounded by broad shoulders.
Some structures are seen in the broad component, particularly on the red side. The narrow component has a width of $\sim$ 3200 km s$^{-1}$ (-2200 to +1000 km s$^{-1}$) and the broad component has an extent of $\sim$ 9000 km s$^{-1}$ (-5000 to + 4000 km s$^{-1}$). The H${\beta}$ profile is somewhat similar, but the broad components are not so pronounced. None of the other lines show a similar structure.
He\,{\sc i} is represented by the 6678 \AA\, line ; the 7065 \AA\ line is weak and the 5876 \AA\ line is absent. He\,{\sc ii} lines at 4686 \AA\ and 5412 \AA\, are strong though we note that the exact strength of the He\,{\sc ii} 4686 \AA\ line is slightly difficult to ascertain as it blended with the N\,{\sc iii} 4640 \AA\ line. The other prominent lines are  O\,{\sc vi} 5292 \AA\,, C\,{\sc iv} 5805 \AA\, and the N\,{\sc iii}  complex at 4640 \AA\, as just mentioned.  In addition, a prominent line is seen at 7727 \AA\ (denoted with a question mark in Figure 6),  which we are unable to identify with
known lines from  spectra of other novae. The interstellar Na D lines have a combined equivalent width of 2.13  \AA\, which translates to a distance of 3.55 kpc according to the empirical formula of Hobbs (1974). This distance estimate  matches well the earlier estimate of 3.3 kpc derived in Section 3.1.
Since the He\,{\sc ii} 4686 \AA\, line is the strongest non-Balmer line in the spectrum, the nova was in the P$_{\rm he^{+}}$ phase according to the Tololo classification (Williams et al 1991, 1994) during this time. Subsequent to this  permitted line phase in early 2001 April, lack of data does not permit us to infer whether V4643 Sgr  progressed through the auroral and nebular stages. But by 2001 mid-June the nova had progressed
to a coronal phase as discussed in the next section.  

\begin{figure}
\resizebox{\hsize}{!}{\includegraphics{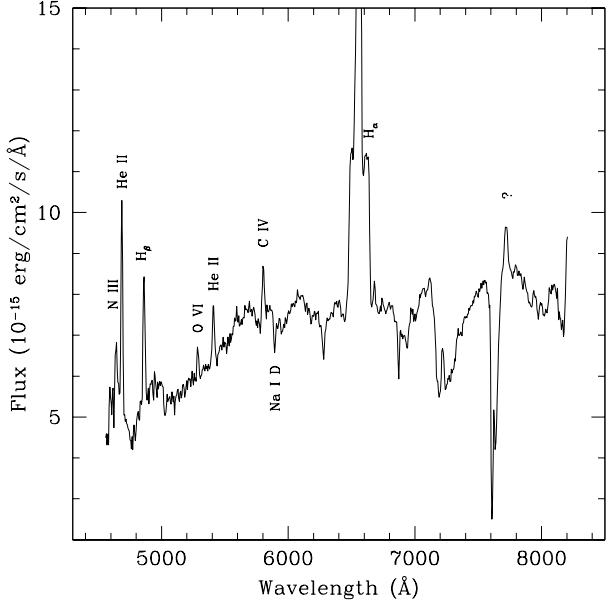}}
\caption[]{The optical spectrum of V4643 Sgr obtained on 2001 April 1 (36 days after outburst) from Vainu Bappu Observatory, India. }
\end{figure}

\begin{figure}
\resizebox{\hsize}{!}{\includegraphics{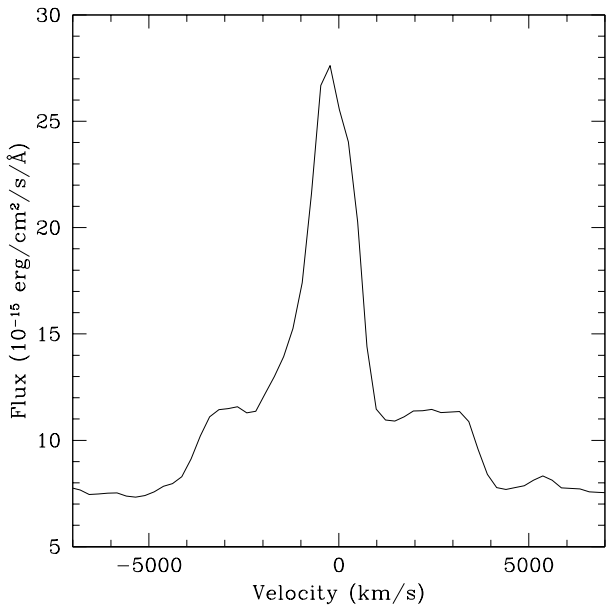}}
\caption[]{H${\alpha}$ emission profile of V4643 Sgr on a velocity scale showing the broad emission flanking the narrow component. }
\end{figure}

\begin{table}
\caption{Line identification in the optical spectrum and observed fluxes relative to H${\beta}$.}
\begin{tabular}{ccc}
\hline
$\lambda$ & Identification & F$_{\lambda}$/F$_{\rm H{\beta}}$ \\
\hline
4640 & N \,{\sc iii} & 0.65 \\
4686 & He \,{\sc ii} & 0.97 \\
5292 & O \,{\sc vi} & 0.19 \\
5412 & He \,{\sc ii} & 0.38 \\
5805 & C \,{\sc iv} & 0.51 \\
6680 & He \,{\sc i} & 0.12 \\
7727 & u.i & 0.43 \\

\hline
F$_{\rm H{\beta}}$ = 1.07$\times$10$^{-13}$ erg/cm$^{2}$/s
\end{tabular}
\end{table}

\subsection{Coronal emission lines}
Figure 4 shows the UKIRT spectra taken on 2001 June 16 in the K band,
112 days after the discovery. The presence of strong [Si\,{\sc vi}] 1.9641 ${\rm{\mu}}$m and [Si\,{\sc vii}] 2.4807 ${\rm{\mu}}$m lines shows that the nova had entered the coronal phase by then. Though  earlier observations of novae show that the coronal phase is observed only after a few hundreds of days, the appearance of strong [Si\,{\sc vi}] and [Si\,{\sc vii}] lines 112 days after discovery is not unexpected considering that V4643 Sgr is a very fast nova with t$_{\rm 3}$= 8.6 days. Previous instances of fast novae exhibiting early occurrence of coronal phase are V838 Her after 17 days and V1500 Cyg after 60 days. The t$_{\rm 3}$ values for these two fast novae are 5 and 3.6 days respectively.  Both the Si lines are broad - with FWZI of 11700 km s${^{\rm -1}}$ and 9850 km s${^{\rm -1}}$ for the [Si\,{\sc vi}] and [Si\,{\sc vii}] lines respectively -  and show considerable structure. Apart from the strong Si lines we additionally detect weaker lines viz. [Ca\,{\sc viii}] at 2.3205 ${\rm{\mu}}$m; a broad structure between 2.03-2.15 ${\rm{\mu}}$m which we attribute to be a blend between 
He\,{\sc i} 2.0581, 2.1120, 2.1132${\rm{\mu}}$m lines; the 2.100 ${\rm{\mu}}$m line which is common in novae but is unidentified (Lynch et al. 2001); and also the He\,{\sc ii} line at 2.1882 ${\rm{\mu}}$m. These weak lines are more clearly seen on enlarging the 2001 June 16 spectrum of Figure 4 but we do not present such a magnified figure here. The [Ca\,{\sc viii}] line is also broad like the Si lines and has a similar FWZI of  10040 km s${^{\rm -1}}$. The K band spectrum taken on 2001 August 12, 138 days since discovery, shows considerable weakening of the [Si\,{\sc vii}] line vis-a-vis that seen in 2001 June. The [Si\,{\sc vi}] line  is still seen clearly but again with reduced strength vis-a-vis the June spectrum (the equivalent widths for the 1.96 and 2.48 ${\rm{\mu}}$m Si lines are 122 \AA\ and 70.5 \AA\ respectively on 2001 August 12).  It is not firmly established, whether the ``coronal" lines in novae arise as a consequence
of  collisional ionization or alternatively from photo-ionization by radiation from the hot central remnant. Based on the compilation by Benjamin $\&$ Dinerstein (1990) of the observed time after outburst when novae have been detected
in the coronal phase, Evans et al. (2003) show that it is likely that
novae enter the coronal phase by a time t${_{\rm cor}}$ $\sim$ (3.34 $\pm$ 1.50)t${_{\rm 3}}$.
For a value of t${_{\rm 3}}$ = 8.6 days for V4643 Sgr (Bruch, 2001), t${_{\rm cor}}$ is estimated to be $\sim$ 28.7  days. This would suggest that at 112 days after
outburst i.e. at the time
of our first detection of the Si coronal lines in V4643 Sgr, the nova was well into the coronal phase. An estimate for the temperature of the hot stellar remnant at such an epoch  can be obtained from the relation T${_{\rm *}}$(t) 
= T${_{\rm 0}}$ exp[0.921(t/t${_{\rm 3}}$)] where T${_{\rm 0}}$ = 15,280K (Bath $\&$ Harkness 1989; Evans et al. 2003). The use of this relation suggests that
T${_{\rm *}}$ was high enough at 112 days to generate the $\geq$ 100 ev photons that
typically characterize the coronal phase (Greenhouse et al. 1990; Evans et al. 2003). Thus we are led to believe,
while allowing for uncertainty in the calculations arising from the above approach, that
the coronal lines in V4643 Sgr are due to photo-ionization. We however note
that there are  other novae where collisional ionization could play
a significant role (Greenhouse et al. 1990; Evans et al. 2003).
 
To summarize, we have presented here spectroscopic and photometric observations of the  fast nova V4643 Sgr which was discovered in outburst in late 2001 February. Our observations, spanning the period between early 2001 March to August, cover the nova's evolution from the early decline stage to the coronal phase. We discuss the behavior and evolution of the prominent emission lines seen in the spectra during this time.

\section*{Acknowledgments}

The research work at Physical Research Laboratory  is funded by the
Department of Space, Government of India.  We thank the UKIRT service
program for observation time in the service mode. UKIRT is operated by
Joint Astronomy Center, Hawaii, on behalf of the UK Particle 
Physics and Astronomy Research Council. We express our thanks to A. Tej for helping with some of the observations and to the VSNET, Japan $\&$ AFOEV, 
France for the use of   optical photometric data from their databases.  We 
thank the referee, Richard Rudy, for his valuable comments which helped improve the manuscript.

\label{lastpage}


\begin{thebibliography}{99}
\bibitem[\protect\citeauthoryear{Ashok}{2001}]{b1} Ashok N.M., Tej A., Banerjee D.P.K., 2001, IAUC 7599
\bibitem[\protect\citeauthoryear{Ashok}{2001}]{b3} Ashok N.M., Banerjee D.P.K., Varricatt W. P., 2001, IAUC 7694
\bibitem[\protect\citeauthoryear{Biese}{2001}]{b5} Ayani K., Kawabata T., 2001, IAUC 7589
\bibitem[\protect\citeauthoryear{Bath}{1989}]{b6} Bath G.T., $\&$ Harkness R.P., 1999, in Classical Novae, ed. M. F. Bode $\&$ A. Evans (Chichester, NY: Wiley), 61
\bibitem[\protect\citeauthoryear{Benjamin}{1990}]{b7} Benjamin R.A., Dinerstein H.L., 1990, AJ, 100, 1588
\bibitem[\protect\citeauthoryear{Bruch}{2001}]{b9} Bruch A., 2001,
IBVS, 5138, 1
\bibitem[\protect\citeauthoryear{chandra}{1992}]{b10} Chandrasekhar T., Ashok N.M., Ragland S., 1992, MNRAS, 255, 412
\bibitem[\protect\citeauthoryear{Della Valle}{2001}]{b11} Della Valle M., Da Silva L., Pompei E., Williams R., 2001, IAUC, 7594
\bibitem[\protect\citeauthoryear{Ennis}{1977}]{b13} Ennis D., Becklin E.E., Beckwith S., Elias J., Gatley I., Mathews K., Neugebauer G., Willner S.P., 1977, ApJ, 214, 478
\bibitem[\protect\citeauthoryear{Evans}{2003}]{b14} Evans A.et al. , 2005, AJ, 126, 1981
\bibitem[\protect\citeauthoryear{Gehrz}{1980}]{b15} Gehrz R.D., 1988, ARA$\&$A, 26, 377
\bibitem[\protect\citeauthoryear{Grandi}{1980}]{b17} Grandi S.A., 1980, ApJ, 238,10
\bibitem[\protect\citeauthoryear{Greenhouse}{1990}]{b19} Greenhouse M.A., 
Grasdalen G.L., Woodward C. E., Benson J., Gehrz R.D., Rosenthal E., $\&$ Skrutskie, M.F., 1990, ApJ, 352,307
\bibitem[\protect\citeauthoryear{Hobbs}{1974}]{b20} Hobbs L.M., 1974, ApJ, 191,381
\bibitem[\protect\citeauthoryear{koornneef}{1985}]{b23} Koornneef J., 1983, A$\&$A, 128, 84
\bibitem[\protect\citeauthoryear{Liller}{2001}]{b25} Liller W., 2001, IAUC, 7589
\bibitem[\protect\citeauthoryear{Lynch et al.}{2000}]{b27} Lynch D.K. 
Rudy R.J., Mazuk S., Puetter R.C., 2000, ApJ, 541, 791 
\bibitem[\protect\citeauthoryear{Lynch et al.}{2001}]{b29} Lynch D.K., 
Rudy R.J., Venturini C., Mazuk S., Puetter R.C., 2001, AJ, 122, 2013 
\bibitem[\protect\citeauthoryear{Nakamura}{2001}]{b31} Nakamura Y., 2001, IAUC, 7591
\bibitem[\protect\citeauthoryear{Samus}{2001}]{b33} Samus N.N., 2001, IAUC, 7591 
\bibitem[\protect\citeauthoryear{Starrfield}{1997}]{b34} Starrfield S., Gehrz R.D., $\&$ Truran J.W., 1997, in AIP Conf. Proc. 402, Astrophysical Implications of the Laboratory  Study of Presolar Materials, ed. T.J. Bernatowicz $\&$ E. K. Zinner (Woodbury: AIP), 203
\bibitem[\protect\citeauthoryear{Storey}{1995}]{b35} Storey P.J., Hummer D.G., 1995, MNRAS, 292, 41
\bibitem[\protect\citeauthoryear{Strittmatter}{1977}]{b37} Strittmatter P.A. et al., 1977, ApJ, 216, 23
\bibitem[\protect\citeauthoryear{van den Bergh}{1987}]{b39} van den Bergh S., Younger P.F., 1987, A$\&$AS, 70, 125
\bibitem[\protect\citeauthoryear{venturini}{2002}]{b41} Venturini S.S., Rudy R.J., Lynch D.K., Mazuk S., Puetter R.C., 2002, AJ, 124, 3009
\bibitem[\protect\citeauthoryear{Williams 1991}{1991}]{b43}Williams R. E., Hamuy M., Phillips M.M., Heathcote S.R., Wells L., Navarrete M., 1991, ApJ, 376, 721 
\bibitem[\protect\citeauthoryear{Williams 1992}{1992}]{b45} Williams R.E., 1992, AJ, 104, 725
\bibitem[\protect\citeauthoryear{Williams 1994}{1994}]{b47}Williams R. E., Phillips M.M., Hamuy M., 1994, ApJS, 90, 297
\end{thebibliography}
\end{document}